\documentclass[aip,rsi,amsmath,amssymb,reprint]{revtex4-1}

\usepackage[english]{babel}

\usepackage[letterpaper,top=2cm,bottom=2cm,left=3cm,right=3cm,marginparwidth=1.75cm]{geometry}
\usepackage{graphicx}
\usepackage{dcolumn}
\usepackage{bm}
\usepackage[utf8]{inputenc}
\usepackage[T1]{fontenc}
\usepackage{mathptmx}
\usepackage{siunitx}

\DeclareSIUnit{\torr}{Torr}

\begin{document}
\pagenumbering{arabic}

\title{A Vacuum-Compatible Cylindrical Inertial Rotation Sensor with Picoradian Sensitivity}

\author{M.~P.~Ross}
\email{mpross2@uw.edu}
\affiliation{Center for Experimental Nuclear Physics and Astrophysics, University of Washington, Seattle, Washington 98195, USA}
\author{J.~van Dongen}
\affiliation{Vrije Universiteit Amsterdam, 1081 HV Amsterdam, Netherlands}
\affiliation{Dutch National Institute for Subatomic Physics, Nikhef, 1098 XG, Amsterdam, Netherlands
}
\author{Y.~Huang}
\author{H.~Zhou}
\author{Y.~Chowdhury}
\author{S.~K.~Apple}
\affiliation{Center for Experimental Nuclear Physics and Astrophysics, University of Washington, Seattle, Washington 98195, USA}
\author{C.~M.~Mow-Lowry}
\author{A.~L.~Mitchell}
\author{N.~A.~Holland}
\affiliation{Vrije Universiteit Amsterdam, 1081 HV Amsterdam, Netherlands}
\affiliation{Dutch National Institute for Subatomic Physics, Nikhef, 1098 XG, Amsterdam, Netherlands
}
\author{B.~Lantz}
\author{E.~Bonilla}
\author{A.~Engl}
\affiliation{Stanford Univserity, Stanford, CA 94305}
\author{A.~Pele}
\author{D.~Griffith}
\author{E.~Sanchez}
\affiliation{California Institute of Technology, Pasadena, CA, 91125, USA}
\author{E.~A.~Shaw}
\author{C.~Gettings}
\author{J.~H.~Gundlach}
\affiliation{Center for Experimental Nuclear Physics and Astrophysics, University of Washington, Seattle, Washington 98195, USA}

\date{\today}

\begin{abstract}
We describe an inertial rotation sensor with a 30-cm cylindrical proof-mass suspended from a pair of 14-{\textmu}m thick BeCu flexures. The angle between the proof-mass and support structure is measured with a pair of homodyne interferometers which achieve a noise level of $\sim 5\ \text{prad}/\sqrt{\text{Hz}}$. The sensor is entirely made of vacuum compatible materials and the center of mass can be adjusted remotely.
\end{abstract}
\maketitle

\section{Introduction}

Sensing minute rotations has long drawn interest from a variety of scientific fields. Recently, rotation sensors with sub-nrad sensitivities have been pursued to improve the seismic isolation systems of gravitational wave observatories \cite{windproofing, diss} and to allow novel measurements of the rotational component of seismic waves \cite{tiltSeis}. 

Multiple devices now reach this sensitivity including ring-laser gyros \cite{ring, ring2, ring3} and flexure-based inertial rotation sensors \cite{BRS, alphra, QRS}. Many of these devices are large (meter-scale) and must be maintained with human intervention making them inadequate for certain applications. Specifically, to operate a sensor within the seismic isolation systems of gravitational wave observatories the sensor must be ultra-high vacuum compatible and remotely operable. 

Here we describe the Cylindrical Rotation Sensor (CRS), a 30-cm scale inertial rotation sensor which reaches a sensitivity of $\sim5\ \text{prad}/\sqrt{\text{Hz}}$ at 1.5 Hz. This design continues our previous sensor development \cite{BRS, cBRS, diss} and shares many qualities with prior designs. The sensor is made of low-outgassing ultra-high-vacuum compatible materials and can be operated and centered remotely. We designed this sensor to improve the rotational seismic isolation performance of gravitational wave observatories. However, we expect the CRS to be applicable in a wide range of research projects, particularly in rotational seismology.

\begin{figure}[!h]
\centering \includegraphics[width=0.5\textwidth]{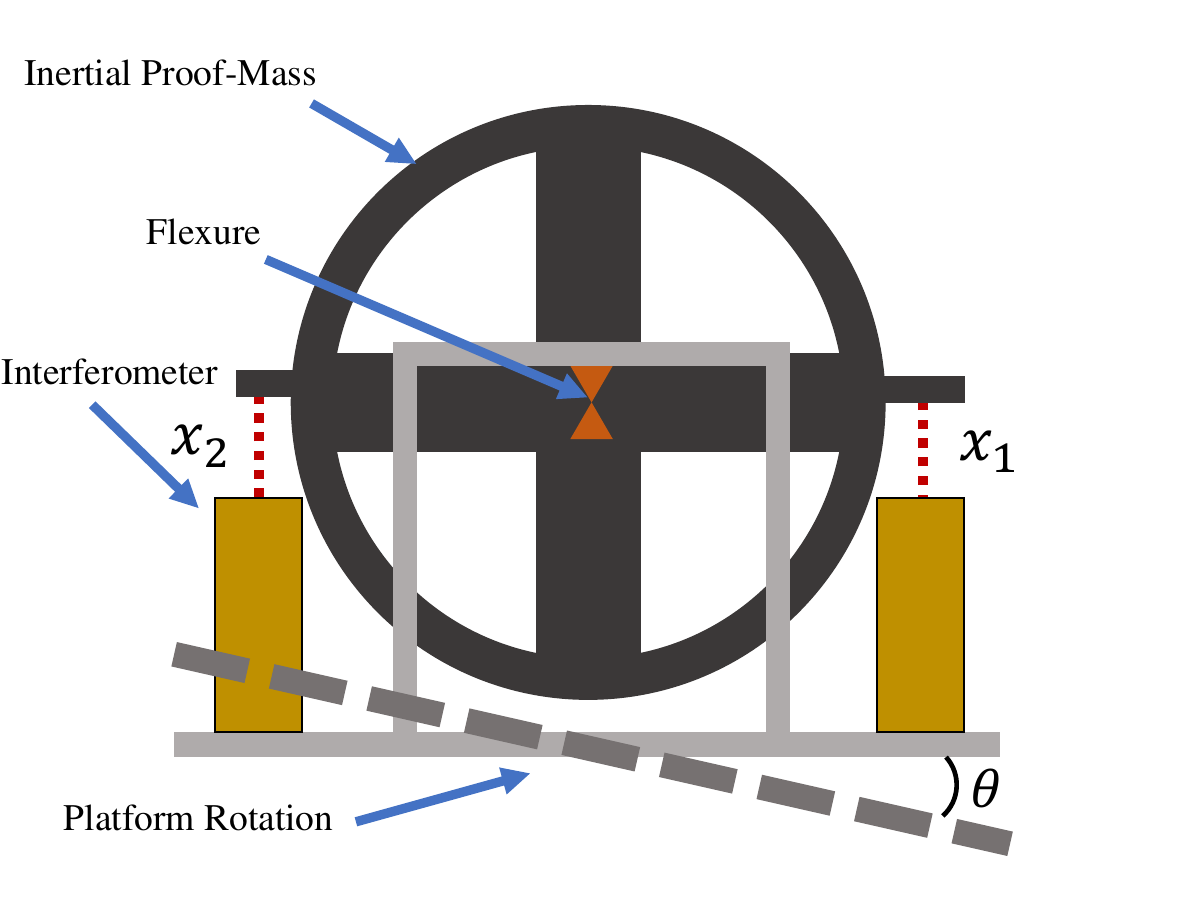}
\caption{A cartoon of the working principle of the CRS. The dark grey indicates the inertial proof-mass that is suspended from the orange flexure, the light grey is the support structure, and the gold rectangles schematically show the interferometers.}
\label{cartoon} 
\end{figure}

\section{Mechanics}

The core mechanism of the CRS is a 30-cm diameter, 5.4-kg aluminum cylindrical proof-mass with a moment of inertia of 0.094 kg-m$^2$ suspended from a pair of 14-{\textmu}m thick BeCu flexures. The center of mass is tuned to be < 22 nm from the pivot point of the flexures corresponding to a translational rejection \cite{diss} of < 1.3 \textmu rad/m. This causes the system to behave as a simple rotational spring-mass system with a resonant frequency of 17 mHz. The proof-mass then acts as an inertial reference above this resonant frequency. The working principle is described by the cartoon shown in Figure \ref{cartoon}. The angle between the support structure and the proof-mass is measured using a pair of homodyne interferometers \cite{HoQI, Cooper_2022}, see Section \ref{hoqi}. As the proof-mass is inertially isolated from motion of the support-structure, angle changes sensed by the readout represent support-structure motion about the axis that runs through the center of the flexures. This allows the device to sense 1-D horizontal angular motion of the surface the sensor is attached to. A detailed description of the dynamics of flexure-based inertial rotation sensors can be found in \citet{diss}.

\begin{figure}[!h]
\centering \includegraphics[width=0.5\textwidth]{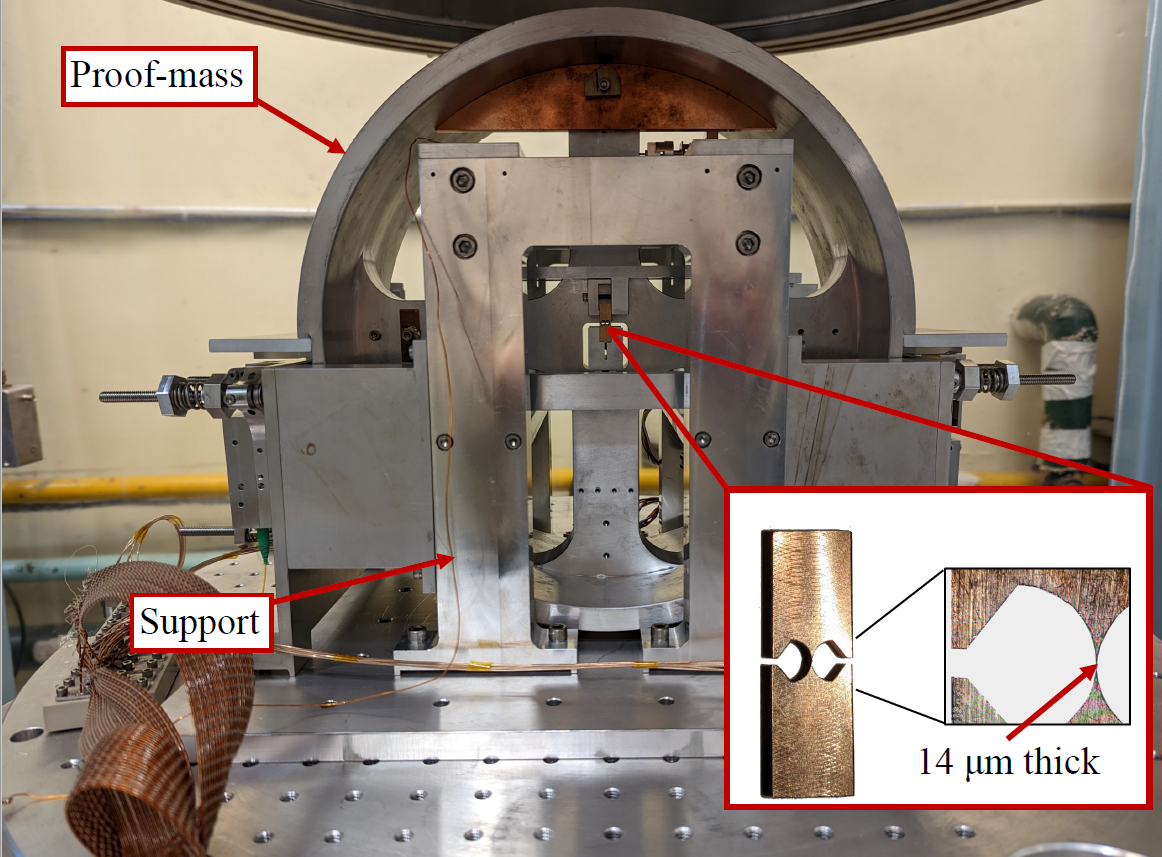}
\caption{Picture of the CRS prototype along with microscope images of the flexures which suspend the proof-mass.}
\label{pic} 
\end{figure}

Figure \ref{pic} shows a picture of the CRS prototype. The proof-mass was machined out of a single monolithic piece of aluminum to maximize thermal uniformity. A seat structure is attached near the center of the cross to which the lower halves of the flexures are mounted with a pair of clamps on either side of the proof-mass. The upper halves of the flexures are mounted to the support structure using similar clamping. The support structure is made of aluminum and is primarily formed by a pair of legs on either side of the proof-mass. These connect through the upper quadrants of the proof-mass. This design increases the stability of the structure. Additionally, the structural pieces are significantly oversized to maximize thermal mass and minimize the impact of high-frequency vibrations.

\section{Readout}\label{hoqi}

\begin{figure}[!h]
\centering \includegraphics[width=0.5\textwidth]{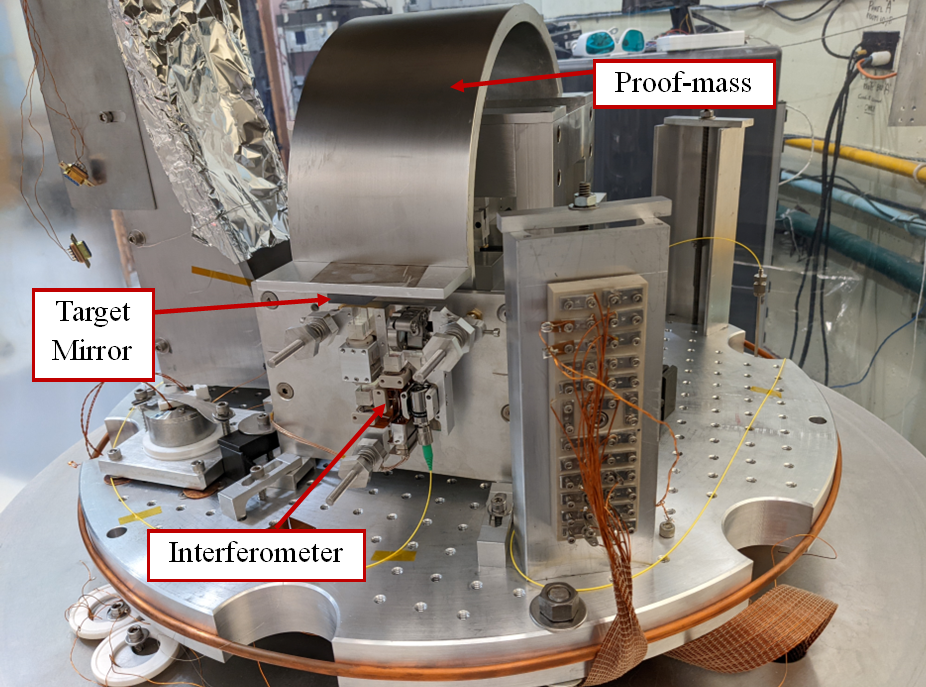}
\caption{Photograph of the homodyne interferometer installed on the side of the sensor. An identical interferometer is installed on the opposite side.}
\label{hoqipic} 
\end{figure}

To significantly improve the rotational performance of gravitational wave observatories, the sensor must outperform the rotational performance of a pair of broadband seismometers located 1-m apart. To meet this requirement, we installed two homodyne interferometers (detailed in \citet{HoQI}) on opposite sides of the proof-mass, shown in Figure \ref{hoqipic}. These deploy a variety of polarization optics to measure multiple phases of the interference pattern produced by a Michelson interferometer. One arm of the interferometer was formed by a mirror attached to the proof-mass, allowing for the distance between the optics and the proof-mass to be measured. The interferometers shared a common laser source (RIO ORION 1064 nm) coupled into the vacuum chamber via fiber optics and split by an in-vacuum fiber splitter. This increased common-mode noise subtraction.

Deploying two interferometers sensing mirrors on either end of the proof-mass allows for the extraction of the angle via:
\begin{equation}
    \theta = \frac{x_1-x_2}{2 r}
\end{equation}
where $x_1$ and $x_2$ are the distance change sensed by the interferometers and $r$= 15.24 cm is the radius from the flexures to each mirror. The differential measurment allows any common noise between the interferometers to be subtracted from the signal of interest. Namely, the frequency noise of the laser that illuminates both interferometers can be minimized.

\section{Remote Centering}

As the CRS was designed to be installed inside the vacuum chambers of gravitational wave observatories, it needed to be operated remotely for long durations. Once suspended, the equilibrium angle of the proof-mass can drift over time due to various physical mechanisms, such as changes in ambient temperature and relaxation of internal stresses. The drifts in the equilibrium angle can drive the proof-mass outside the range of the interferometer and even cause it to rest on its mechanical stops. The angular range of the interferometers is $\sim$ mrad and is limited by beam spot displacement due to the angle of the proof-mass mirrors.

To counter this drift, we can shift the proof-mass's horizontal center-of-mass using the sensor's remote mass adjuster. This process is temporarily disruptive to the sensor's performance yet is only needed occasionally. Some commercial broadband seismometers have a similar centering mechanism.

The remote mass adjuster consists of a 1-gram brass mass attached to a 0-90 screw that is allowed to rotate but is held in place by a BeCu leaf spring. One edge of the mass is in contact with a flat which allows the mass to be precisely translated by rotating the screw. This assembly is installed on the cross of the proof-mass. Details of the centering mechanism can be found in \citet{diss}.

Running wires to the proof-mass to power a motor would be significantly stiffer than the flexures and ruin the performance of the sensor. To alleviate this issue, when adjustment is needed, a motor attached to the support structure turns the screw. A set of claws with intentionally large backlash couple the motor to the adjuster. This coupling allows the motor to rotate the adjuster while making contact, then back rotate to mechanically decouple. Once decoupled, the sensor returns to its previous dynamics with a shifted equilibrium angle. 

\begin{figure}[!h]
\centering \includegraphics[width=0.5\textwidth]{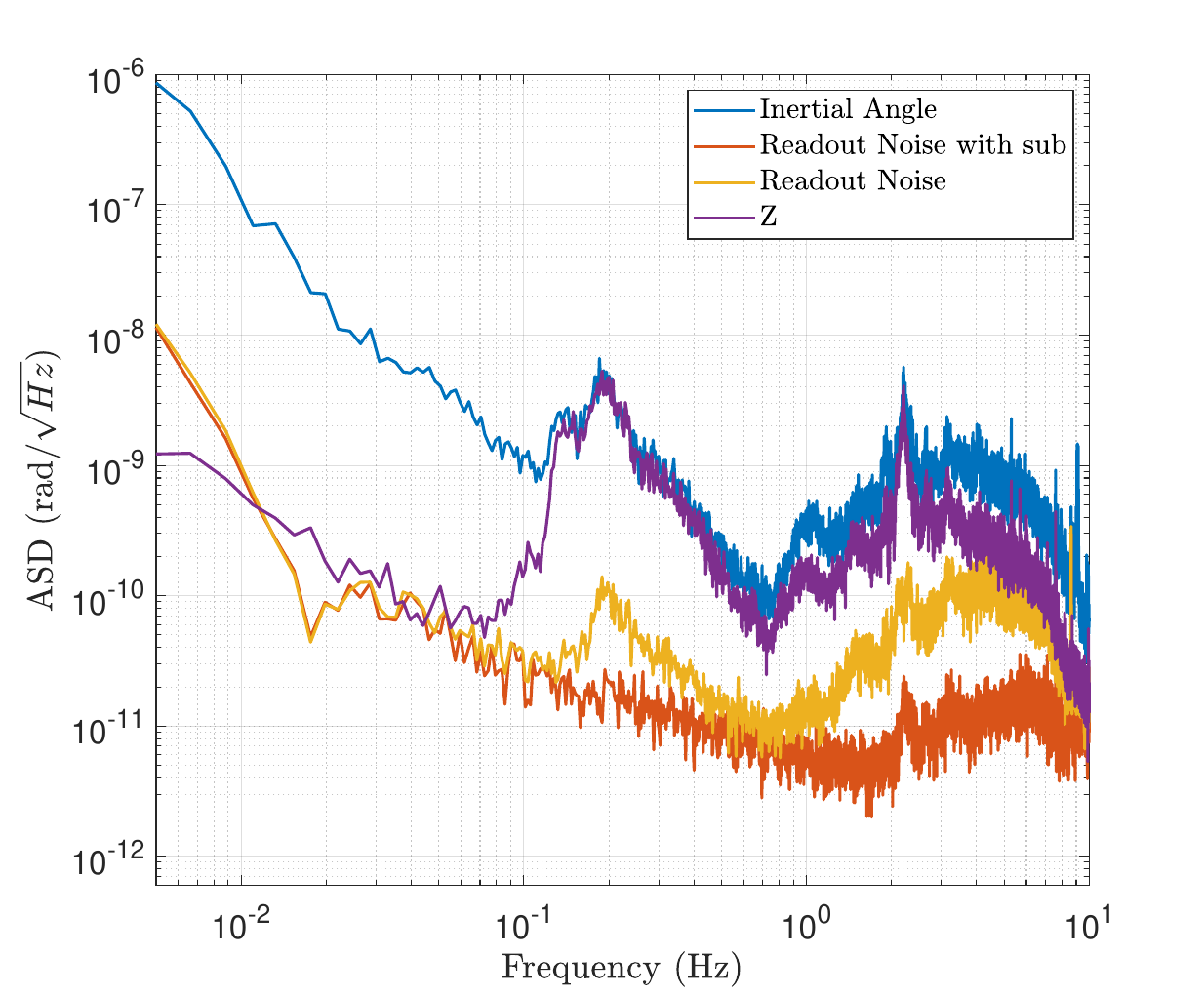}
\caption{Amplitude spectral density displaying the inertial angle, readout noise estimate with seismometer subtraction, readout noise without seismometer subtraction, and the vertical component of the seismometer in arbitrary units. All signals have been corrected for the respective sensor response. The horizontal seismometer channels are omitted for clarity.}
\label{sub} 
\end{figure}

\section{Noise Performance}

The CRS was tested in a bell-jar vacuum chamber housed in a defunct-cyclotron cave at the Center for Experimental Nuclear Physics and Astrophysics on the campus of University of Washington. The cave provided thermal stability but the location had a high level of seismic activity as it was on an urban campus near multiple high-traffic roads. 

To assess the intrinsic noise of the instrument, we calculated the residuals of a coherent subtraction between the two readouts. This subtraction is conducted with the {\tt mccs2} algorithm \cite{mccs2}, which removes the coherent part of two signals to leave only the incoherent noise. For the CRS, this represents the combined noise contribution of the two readouts and is plotted in Figure \ref{sub} along with the observed angle. 

We found that ambient seismic motion was coupling into the readout noise measurements through vibrations of the fiber optics. To assess the expected performance of the sensor in a quiet seismic environment (i.e. a seismic isolation platform), we attached a MBB-2 \cite{} three-axis seismometer to the vacuum chamber. The three seismometer channels were added into the coherent subtraction to remove this spurious coupling from the readout noise estimations. The readout noise with and without this additional subtraction is shown in Figure \ref{sub} along with the vertical axis of the seismometer. The seismometer subtraction removes excess noise mainly above 1 Hz and at the microseism (0.2 Hz). The residual readout noise reaches a maximum sensitivity of $\sim\ 5\ \text{prad}/\sqrt{\text{Hz}}$ at 1.5 Hz. This noise level is a factor of 50 improvement over our group's previous devices \cite{diss}. The improved performance is primarily due to the deployment of high sensitivity interferometer readouts\cite{HoQI} with minor improvements at low-frequencies due to the monolithic cylindrical test mass.

\begin{figure}[!h]
\centering \includegraphics[width=0.5\textwidth]{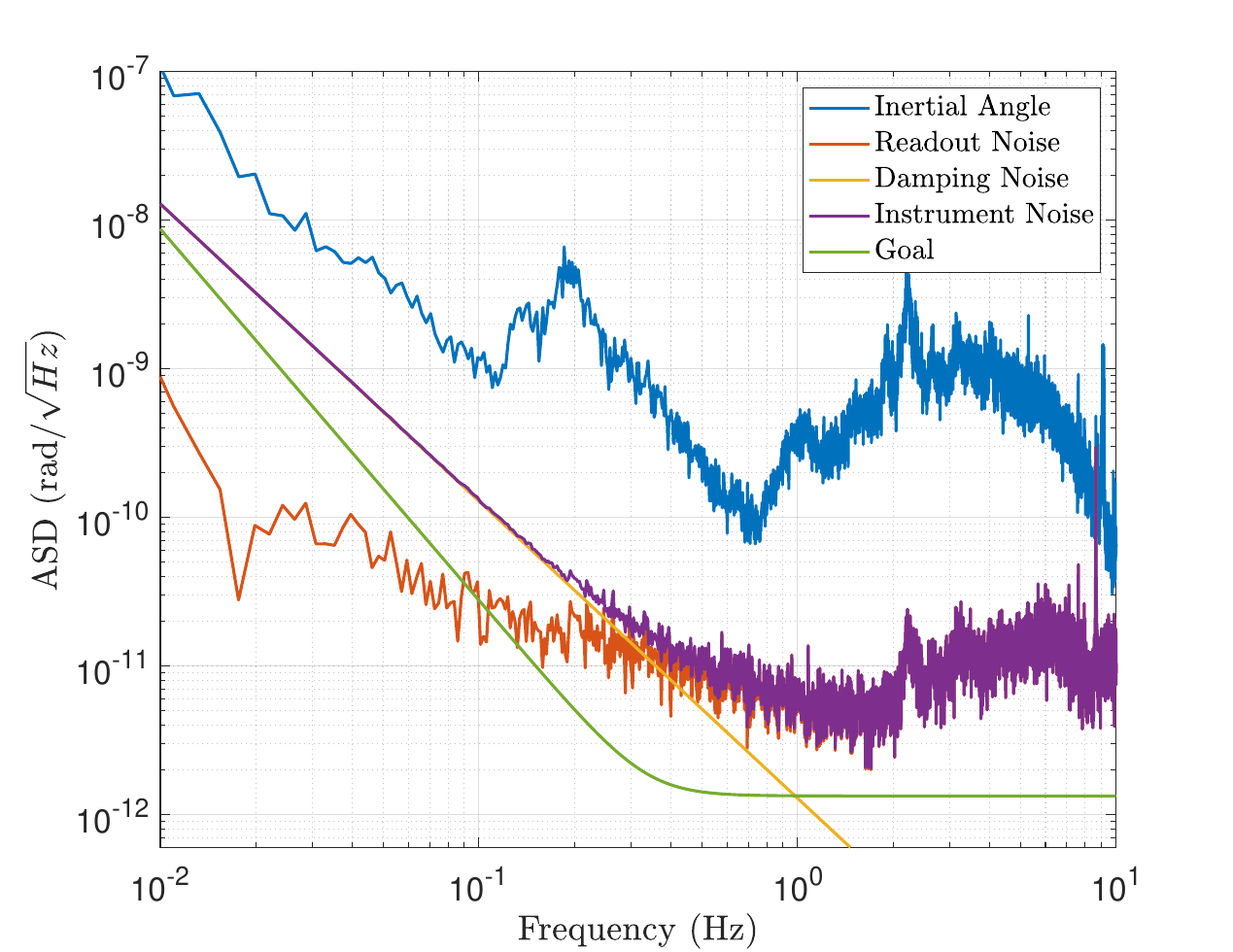}
\caption{Amplitude spectral density of the inertial angle, readout noise, external damping noise estimate, the total instrument noise, and the sensor performance goal.}
\label{asd} 
\end{figure}

Readout noise is not the only noise source that can limit inertial rotation sensors. Any effect that changes the angle of the proof-mass is indistinguishable from rotations of the platform. Accurately assessing these contributions is difficult with a single sensor. However, some fundamental noise sources can be calculated from first principles. The residual pressure of the current vacuum chamber can only reach $\sim$70~\textmu~Torr. Thus, damping due to residual pressure dominates the mechanical loss of the sensor. Figure \ref{asd} shows the residual pressure damping noise \cite{thermal} calculated for an observed quality factor of 294. The damping noise limits the performance of the sensor below 0.2 Hz with the readout dominating above that.

We believe the sensor noise is well represented above 0.1 Hz with the combination of damping and readout noise shown in Figure \ref{asd}. The observed angle in this frequency band is then the angular component of the ambient seismic wavefield with the peak at $\sim$ 0.2 Hz being the oceanic microseism and the rise above 1 Hz being anthropogenically sourced. 

Also shown in Figure \ref{asd} is the final noise goal of the instrument. With improvements to the vacuum, the sensor will be limited by internal losses instead of external damping. This is expected to improve the observed quality factor from 294 to > 1000. Additionally, we expect further reduction in the readout noise when the sensor is deployed in a seismically quiet environment. 

\section{Conclusion}

We have constructed an interferometrically readout inertial rotation sensor with a cylindrical proof-mass, which achieves $< \text{nrad}/\sqrt{\text{Hz}}$ noise above 35 mHz and reaches a maximum sensitivity of $\sim 5\ \text{prad}/\sqrt{\text{Hz}}$ at 1.5 Hz. This sensor is vacuum compatible and allows for remote mass centering. We plan further sensitivity improvements in the near future. With these, the sensor is expected to have a three-fold enhancement in sensitivity as compared to the current prototype.

Similar sensors will soon be installed at the LIGO gravitational-wave observatories which will significantly improve the observatories' seismic isolation. Additionally, the sensor's applications to seismology are actively being explored.

\section{Data Availability Statement}

Schematics of the CRS can be found at \url{https://github.com/mpross/CRS-Schematics}. Code and data to generate the plots shown here can be found at \url{https://github.com/mpross/CRS-Analysis}.

\section{Acknowledgements}

Participation from the University of Washington, Seattle, was supported by funding from the NSF under Awards PHY-1607385, PHY-1607391, PHY-1912380, and PHY-1912514. This project has received funding from the European Research Council (ERC) under the European Union's Horizon 2020 research and innovation programme (grant agreement No. 865816).

\bibliography{CRS.bib}

\end{document}